\documentclass[11pt]{article}

\usepackage[utf8]{inputenc}
\usepackage[T1]{fontenc}
\usepackage[margin=1in]{geometry}
\usepackage{amsmath,amssymb,amsfonts}
\usepackage{graphicx}
\graphicspath{{figs/}{./}}
\usepackage{booktabs}
\usepackage{float}
\usepackage{multirow}
\usepackage{array}
\usepackage{siunitx}
\usepackage{xcolor}
\usepackage{subcaption}
\usepackage{enumitem}
\usepackage[numbers,sort&compress]{natbib}
\usepackage[hidelinks]{hyperref}
\usepackage{cleveref}

\usepackage{tikz}
\usetikzlibrary{positioning,arrows.meta,shapes.geometric,fit,backgrounds,calc}
\usepackage{pgfplots}
\pgfplotsset{compat=1.18}

\newcommand{\Sys}{FIDES}
\newcommand{\concept}{concordance}
\newcommand{\Concept}{Concordance}
\newcommand{\saydo}{\textsf{say}\ensuremath{\rightarrow}\textsf{do}}
\newcommand{\doreal}{\textsf{do}\ensuremath{\rightarrow}\textsf{real}}
\newcommand{\sayresult}{\textsf{say}\ensuremath{\rightarrow}\textsf{result}}
\newcommand{\bh}{buy-and-hold}
\newcommand{\Bh}{Buy-and-hold}
\newcommand{\strat}{\texttt{strategy(df)}}

\definecolor{saycol}{RGB}{250,219,178}
\definecolor{docol}{RGB}{197,220,246}
\definecolor{realcol}{RGB}{198,232,197}
\definecolor{gatered}{RGB}{247,199,199}
\definecolor{edgegray}{RGB}{90,90,90}
\definecolor{c41mini}{RGB}{200,50,50}
\definecolor{c4o}{RGB}{40,90,180}
\definecolor{cqwen}{RGB}{30,140,110}
\definecolor{c4omini}{RGB}{230,150,25}
\definecolor{c2stage}{RGB}{140,80,180}

\title{\Sys{}: A \Concept{} Protocol for LLM-Generated Trading Strategies}
\author{
  Arther Tian\textsuperscript{a},
  Alex Ding\textsuperscript{a,*},
  Simon Wu\textsuperscript{a},
  Aaron Chan\textsuperscript{a}\\
  \textsuperscript{a}DGrid AI\\[0.5em]
  \textsuperscript{*}Corresponding author: \texttt{alex.ding@dgrid.ai}
}
\date{}

\begin{document}
\maketitle

\begin{abstract}
An LLM asked for a trading strategy returns three artifacts at once: a
natural-language rationale, an executable implementation, and---once
run---a track record. Whether these are the same object is rarely checked.
We present \Sys{}, a measurement protocol that treats them as three views
to be reconciled rather than one deliverable to be graded. Through
\emph{dual delivery}---a single model call that returns both a
natural-language strategy (with an explicit claimed edge) and a
self-contained \strat{} function---\Sys{} executes the code in a sandbox
against a lag-one out-of-sample backtest the model cannot game, and scores
three \concept{} gaps in $[0,1]$: \saydo{} (does the code implement the
stated rules?), \doreal{} (does it execute as written, without
look-ahead?), and \sayresult{} (does the claimed edge survive the
numbers?). On 8 liquid US ETFs across four models plus a two-stage
elicitation arm (40 strategies; 2023--24 out-of-sample), three findings
stand out, and none is flattering. First, \concept{} does not predict
profit: the most speech--code consistent model (\saydo{} $=1.00$) still
loses to \bh{}, only $2$ of $40$ strategies beat it, and a plain
\textsc{sma}(50,200) rule outperforms every model's mean Sharpe. Second,
self-assessment is badly calibrated: $32$ of $40$ strategies claim to beat
\bh{} and exactly one does---a $3.1\%$ hit rate. Third, the language--code
judge is not an oracle: swapping it for a second model flips \saydo{} on
more than half of items (exact agreement $16/40$, mean $|\Delta|=0.15$). A
look-ahead dose--response validates the execution gap---injecting
\texttt{Close.shift(-1)} drops \doreal{} by $0.33$ on average---while we
disclose, as a negative result, that our runtime future-information probe
fired on neither clean nor injected code. We frame \Sys{} as a protocol for
measurement fidelity, not a claim about market performance, and delimit
external validity---including a model roster constrained to what our
inference catalog served---as future work.
\end{abstract}

\section{Introduction}
\label{sec:intro}

Large language models are increasingly asked not merely to discuss markets
but to produce trading strategies outright---as natural-language
rationales, as runnable code, and, through agent frameworks, as systems
that place orders~\citep{lopezlira2023chatgpt,xiao2024tradingagents,yu2023finmem}.
A single such request returns three artifacts at once: the strategy the
model \emph{states} in prose, the strategy its \emph{code} actually
implements, and the track record that strategy \emph{realizes} when run. It
is natural to treat these as one deliverable and grade it---did the code
run, was the return positive---but that skips the prior question of whether
the three artifacts even describe the same strategy. A fluent rationale can
be paired with code that quietly does something else; code that runs
cleanly can be silently peeking at future bars; and a confident claim to
beat the market can sit atop a track record that does not. In finance these
gaps are not cosmetic: acting on a stated rationale when the code diverges,
or trusting a backtest that looked ahead, loses real money.

No existing evaluation reconciles the three views jointly. Code-generation
benchmarks measure functional correctness against hidden unit
tests~\citep{chen2021humaneval,austin2021mbpp}, but there is no unit test
for \emph{did the code implement the English}, and a generated trading
strategy has no reference implementation to diff against. Financial LLM
benchmarks grade domain knowledge and question
answering~\citep{wu2023bloomberggpt,xie2023pixiu,xie2024finben,yang2023fingpt},
not the internal consistency of a produced strategy. Return-prediction and
trading-agent studies pursue profit
directly~\citep{lopezlira2023chatgpt,xiao2024tradingagents,yu2023finmem} and
thereby inherit the well-documented hazards of that target---backtest
overfitting, data snooping, and look-ahead
bias~\citep{bailey2014pseudomath,bailey2017pbo,white2000reality,sullivan1999datasnooping}---hazards
acute enough that beating a passive benchmark out of sample is the
exception, not the rule~\citep{fama1970efficient,bailey2014deflated}, and
severe enough that look-ahead leakage is now studied specifically for
LLMs~\citep{benhenda2026lookahead}. Closest in spirit, work on
chain-of-thought faithfulness shows that a model's stated reasoning need not
reflect the computation that produced its
answer~\citep{turpin2023unfaithful,lanham2023faithfulness}; but it operates
on multiple-choice reasoning, where the ``answer'' is a label, rather than
on an artifact whose correctness is settled by a non-gameable outcome.

\begin{figure}[t]
  \centering
  \resizebox{0.92\linewidth}{!}{%
  \begin{tikzpicture}[
    font=\small,
    >={Stealth[length=2.6mm]},
    art/.style={draw=edgegray,rounded corners=3pt,align=center,
                inner sep=6pt,minimum height=13mm},
    gap/.style={->,draw=edgegray,line width=0.9pt},
    glbl/.style={align=center,font=\footnotesize,fill=white,inner sep=2pt},
  ]
    \node[art,fill=saycol,minimum width=46mm] (say) at (0,4.5)
      {\textbf{What the model \emph{says}}\\[1pt]
       \footnotesize natural-language strategy\\
       \footnotesize $+$ claimed edge};
    \node[art,fill=docol,minimum width=44mm] (do) at (-5.2,0)
      {\textbf{What it \emph{does}}\\[1pt]
       \footnotesize generated \strat{}\\
       \footnotesize $\rightarrow$ signals $\in\{-1,0,1\}$};
    \node[art,fill=realcol,minimum width=50mm] (real) at (5.2,0)
      {\textbf{What \emph{really happens}}\\[1pt]
       \footnotesize lag-1 out-of-sample backtest\\
       \footnotesize \textbf{non-gameable ground truth}};
    \draw[gap] (say) -- (do);
    \draw[gap] (say) -- (real);
    \draw[gap] (do) -- (real);
    \node[glbl,align=right,anchor=east]
      at ($(say)!0.55!(do)+(-1.5mm,0)$) {\textbf{\saydo{}}\\ NL vs.\ code (fixed judge)};
    \node[glbl,align=left,anchor=west]
      at ($(say)!0.55!(real)+(1.5mm,0)$) {\textbf{\sayresult{}}\\ claimed edge vs.\ OOS numbers};
    \node[glbl,align=center,anchor=north]
      at ($(do)!0.5!(real)+(0,-1.2mm)$) {\textbf{\doreal{}}\\ executes as written, no look-ahead};
  \end{tikzpicture}}
  \caption{The three artifacts of an LLM trading strategy and the three gaps
  \Sys{} measures. A single model call (dual delivery) yields a
  natural-language strategy and its code; executing the code on a lag-one
  out-of-sample backtest yields a realized track record. Only the backtest
  is non-gameable ground truth. Each gap is scored in $[0,1]$ ($1=$ fully
  concordant); profit is treated as a side metric, not the objective.}
  \label{fig:triangle}
\end{figure}
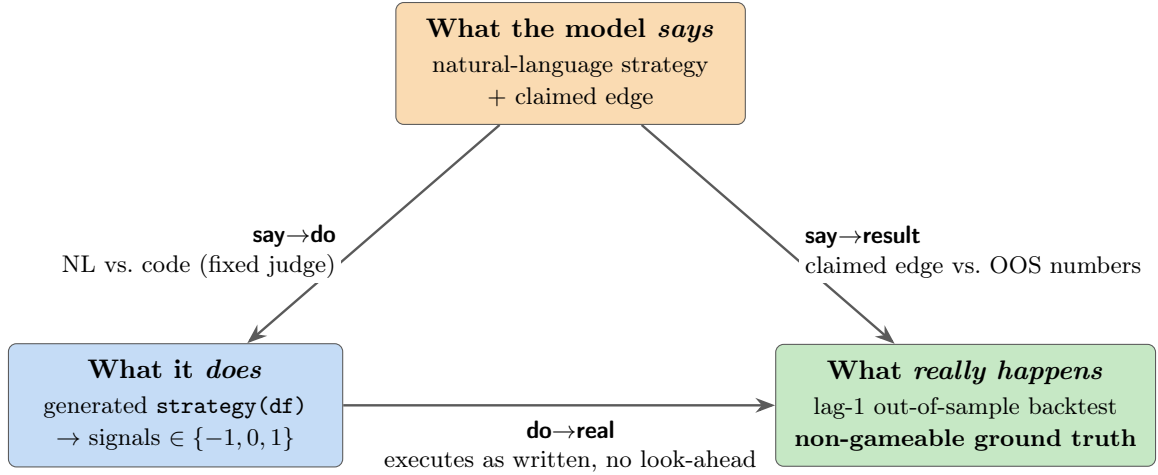

We introduce \Sys{}, a measurement protocol that reconciles the three
artifacts rather than grading one. \Sys{} uses \emph{dual delivery}: a
single model call returns both a natural-language strategy (with an explicit
claimed edge) and a self-contained \strat{} function, so the language and
the code come from one act of reasoning and can be compared without a
translation step. It then executes the code in a sandbox and scores three
\concept{} gaps, each in $[0,1]$ with $1$ fully concordant: \saydo{} asks
whether the code implements the stated rules, scored by a fixed,
non-candidate language--code judge; \doreal{} asks whether the code executes
as written, without static look-ahead; and \sayresult{} asks whether the
claimed edge survives the out-of-sample numbers. The load-bearing design
choice is the anchor: outcomes come from a lag-one out-of-sample backtest
the model cannot game---positions act on the prior bar, so the code cannot
trade on information it should not have---and \doreal{} and \sayresult{} are
thereby grounded in what actually happened rather than in another model's
opinion (\Cref{fig:triangle}). Profit is a side metric, and we deliberately
do not repair or search over strategies to raise it: doing so would
optimize the very quantity whose relationship to the stated logic we are
trying to measure.

Across 8 liquid US ETFs and four models (40 strategies; 2023--24
out-of-sample), the protocol turns up three results, none of them flattering
to the models. \Concept{} does not predict profit: the most speech--code
consistent model scores a perfect \saydo{} of $1.00$ yet loses to \bh{},
only $2$ of $40$ strategies beat it, and a plain \textsc{sma}(50,200) rule
outperforms every model's mean Sharpe. Self-assessment is badly calibrated:
$32$ of $40$ strategies claim to beat \bh{} and exactly one does, a $3.1\%$
hit rate. And the language--code judge is not an oracle---swapping it for a
second model flips \saydo{} on more than half of items (exact agreement
$16/40$, mean $|\Delta|=0.15$), so a single judge's scores should be read as
one measurement, not ground truth. We validate the execution gap with a
look-ahead dose--response (injecting \texttt{Close.shift(-1)} drops
\doreal{} by $0.33$ on average) and, in the same spirit, disclose a negative
result: our runtime future-information probe fired on neither clean nor
injected code---a limitation of that probe rather than evidence of clean
execution. This cheap-first, measure-what-you-can, audit-the-judge stance is
shared with a broader line of the authors' work on trustworthy
evaluation~\citep{tian2025costaware,tian2026multidim,tian2026poqjudge,tian2026sfga};
here the object---an LLM's own trading strategy---and the mechanism---three
gaps anchored by a non-gameable backtest---are new. We present \Sys{} as a
protocol for measurement fidelity, not as a claim about market performance,
and we are explicit that the model roster is constrained by what our
inference catalog actually served (\Cref{sec:setup}).

\Sys{} makes three contributions:
\begin{itemize}[leftmargin=1.4em,itemsep=2pt]
  \item \textbf{A \concept{} protocol for generated strategies.} Dual
  delivery plus three interval-valued gaps---\saydo{}, \doreal{},
  \sayresult{}---that reconcile an LLM's stated logic, its code, and its
  realized track record, anchored by a lag-one out-of-sample backtest the
  model cannot game (\Cref{sec:design}).
  \item \textbf{An empirical study with uncomfortable findings.} On 8 ETFs
  across four models we show that \concept{} does not predict profit, that
  models' self-reported edge is badly calibrated, and that the
  language--code judge is not robust to being swapped---each quantified
  against passive, rule-based, and cost-adjusted baselines
  (\Cref{sec:results}).
  \item \textbf{Honest diagnostics and delimited scope.} A validated
  look-ahead dose--response, a disclosed failed runtime probe, a zero-LLM
  rule baseline, and a two-stage elicitation control, framed throughout as
  measurement fidelity with external validity---including model
  coverage---delimited as future work (\Cref{sec:results,sec:discussion}).
\end{itemize}

\section{Related Work}
\label{sec:related}

\paragraph{LLMs for finance and trading.}
Domain models such as BloombergGPT~\citep{wu2023bloomberggpt} and open
efforts such as FinGPT~\citep{yang2023fingpt}, together with benchmark
suites~\citep{xie2023pixiu,xie2024finben}, establish that LLMs can be
adapted to financial text; return-prediction and agentic-trading studies go
further and ask models to forecast or to
trade~\citep{lopezlira2023chatgpt,xiao2024tradingagents,yu2023finmem}. These
efforts grade domain knowledge or chase realized profit. Neither asks our
question---whether a single generated strategy's prose, code, and outcome
agree---which is prior to, and orthogonal to, whether the strategy makes
money.

\paragraph{Evaluating generated code.}
Functional-correctness benchmarks score generated programs against held-out
unit tests~\citep{chen2021humaneval,austin2021mbpp}, treating execution as
ground truth. We adopt the same execute-to-verify philosophy but face a
harder object: a trading strategy has no reference implementation, so there
is no test that says whether the code matches the stated rules. \Sys{}
therefore pairs execution (for \doreal{} and \sayresult{}) with an explicit
language--code comparison (for \saydo{}), and anchors the former in a
non-gameable market backtest rather than a unit test.

\paragraph{Faithfulness of stated reasoning.}
A model's stated rationale need not reflect the computation that produced
its output: chain-of-thought explanations can be plausible yet
unfaithful~\citep{turpin2023unfaithful,lanham2023faithfulness}, even as
chain-of-thought and self-consistency improve accuracy on
reasoning~\citep{wei2022cot,wang2022selfconsistency}. This literature
motivates \saydo{}, but it operates where the ``answer'' is a label; \Sys{}
moves the question to an artifact---code---whose behaviour is settled by a
backtest, so the say--do gap can be checked against something the model
cannot argue with.

\paragraph{LLM-as-judge and its biases.}
Strong models are now standard evaluators~\citep{zheng2023mtbench,liu2023geval},
but their verdicts carry documented biases---order effects, verbosity, and
self-preference---and shift with superficial
factors~\citep{wang2024notfair,chen2024judgebias}. Because \saydo{} is the
one gap we score with an LLM, we treat the judge as a measurement to be
audited: we fix a non-candidate judge and re-score every item with a second
model (\Cref{sec:results}), reporting the disagreement as a first-class
result rather than assuming the judge is an oracle.

\paragraph{Backtest overfitting, data snooping, and look-ahead bias.}
A large finance literature shows that impressive backtests routinely fail
out of sample---through overfitting, selection, and multiple
testing~\citep{bailey2014pseudomath,bailey2017pbo,bailey2014deflated}, and
through data snooping over trading rules~\citep{white2000reality,sullivan1999datasnooping}---so
that beating a passive benchmark is rare~\citep{fama1970efficient}, and
Sharpe ratios must be read with these hazards in mind~\citep{sharpe1994ratio}.
Look-ahead leakage in particular is now studied for
LLMs~\citep{benhenda2026lookahead}. We take these as design constraints: our
outcome anchor is a strict lag-one out-of-sample backtest, profit is never
optimized inside the loop, and \doreal{} includes an explicit look-ahead
check whose sensitivity we validate by injection.

\paragraph{Trustworthy, cost-aware evaluation.}
\Sys{} continues a line of the authors' work on cheap-first, self-auditing
evaluation~\citep{tian2025costaware,tian2026adaptive,tian2026multidim,tian2026poqjudge,tian2026sfga}.
The shared stance---measure what cheap signals can settle, escalate or
distrust only where they cannot, and quantify the fallible component rather
than hide it---carries over; the object (an LLM's own trading strategy) and
the mechanism (three gaps anchored by a non-gameable backtest) are new.

\section{Protocol Design}
\label{sec:design}

\subsection{Problem setup and notation}
A \emph{task} fixes a ticker and two date windows: an in-sample window
(shown to the model as context only---no in-sample fitting is performed) and
an out-of-sample (OOS) window on which the strategy is scored. A single
model call on a task yields three things: a natural-language strategy $s$, a
self-contained function $c=\strat{}$, and a one-sentence claimed edge $e$.
Executing $c$ produces a signal series $\sigma\in\{-1,0,1\}^{T}$ (short,
flat, long), which a fixed backtester turns into an OOS Sharpe $S$ and
return $R$, alongside \bh{} references $S_{\text{bh}},R_{\text{bh}}$. \Sys{}
reports three gaps, each in $[0,1]$ with $1$ concordant:
$g_{\saydo{}}(s,c)$, $g_{\doreal{}}(c,\sigma)$, and
$g_{\sayresult{}}(e,S,R)$. Profit ($S,R$) is retained as a side metric; the
gaps, not the profit, are the object of study.

\subsection{Dual delivery}
The central elicitation choice is to obtain the prose and the code from
\emph{one} completion. A single prompt requests a rule-based daily strategy
(inputs, entry/exit, risk controls, and a line \texttt{Claimed edge: ...})
and a function \strat{} returning a signal series, under the explicit
instruction to use only information available at the close of day $t$ to
trade $t{+}1$. Because both artifacts come from one act of reasoning, the
say--do comparison is not confounded by a separate translation step. As a
control we also run a \emph{two-stage} variant that elicits the prose and
then, in a second call, asks for code implementing that prose; we treat it
as an elicitation ablation, not as a distinct model (\Cref{sec:results}).

\subsection{The three gaps}
\paragraph{\saydo{}: language vs.\ code.}
A fixed judge model $J_0$---deliberately not one of the models under
test---reads the prose and the code and returns four binary checks: whether
the long/short/flat \emph{direction} rules match ($d$), whether the
\emph{indicators and lookbacks} named in the text appear in the code ($i$),
whether \emph{exits and risk} rules match or are jointly absent ($x$), and
whether the code adds \emph{no material extra rule} ($m$). The gap averages
them,
\begin{equation}
  g_{\saydo{}} \;=\; \tfrac{1}{4}\,(d+i+x+m), \qquad d,i,x,m\in\{0,1\},
  \label{eq:saydo}
\end{equation}
scored at temperature $0$. As a zero-LLM ceiling we also compute a rule
variant $g^{\text{rule}}_{\saydo{}}$: the fraction of lookback numbers named
in the prose that also appear literally in the code.

\paragraph{\doreal{}: code vs.\ execution.}
Executed against real OHLCV in a sandbox, the code earns three binary
credits---it \emph{runs} to a valid signal series ($\rho$); the signals are
\emph{not constant} ($\alpha$, i.e.\ not a degenerate all-long/short/flat
series); and the code contains \emph{no static look-ahead} ($\ell$), where
$\ell{=}0$ if any of a small set of leakage patterns (e.g.\
\texttt{.shift(-n)}, \texttt{.pct\_change(-n)}, \texttt{center=True},
\texttt{.iloc[...+1]}) match the source:
\begin{equation}
  g_{\doreal{}} \;=\; \tfrac{1}{3}\,(\rho+\alpha+\ell), \qquad
  \rho,\alpha,\ell\in\{0,1\}.
  \label{eq:doreal}
\end{equation}

\paragraph{\sayresult{}: claim vs.\ outcome.}
The claimed edge is parsed for testable assertions and each is checked
against the OOS numbers: a claim to \emph{beat} \bh{} requires
$S>S_{\text{bh}}\wedge R>R_{\text{bh}}$; a claim about \emph{Sharpe}
requires $S>0$; a claim of \emph{profit} requires $R>0$. The gap is $1$ if
all triggered checks pass, $0$ if any fails, and $0.5$ if no testable claim
is detected:
\begin{equation}
  g_{\sayresult{}} =
  \begin{cases}
    1 & \text{all triggered checks pass},\\
    0 & \text{some triggered check fails},\\
    0.5 & \text{no testable claim detected.}
  \end{cases}
  \label{eq:sayresult}
\end{equation}

\subsection{The non-gameable backtest}
Outcomes come from a strict lag-one engine: the position on day $t$ is the
signal from day $t{-}1$, so a strategy cannot act on the same-day close it
is reacting to,
\begin{equation}
  \text{pos}_t = \sigma_{t-1}, \qquad
  r_t = \text{pos}_t\!\left(\tfrac{C_t}{C_{t-1}}-1\right)
        - \tfrac{b}{10^{4}}\,\lvert \text{pos}_t-\text{pos}_{t-1}\rvert,
  \label{eq:backtest}
\end{equation}
with annualized Sharpe $S=\sqrt{252}\,\overline{r}/\mathrm{std}(r)$. The main
run uses zero cost ($b{=}0$); a robustness ablation uses $b{=}10$ bp one-way
on turnover. \Bh{} sets $\text{pos}_t\equiv 1$ on the same OOS slice; a
long-only \textsc{sma}(50,200) crossover on the same engine is a non-LLM
rule baseline. Strategies are never repaired or searched to raise $S$, which
would break the identification the protocol provides.

\subsection{Sandbox}
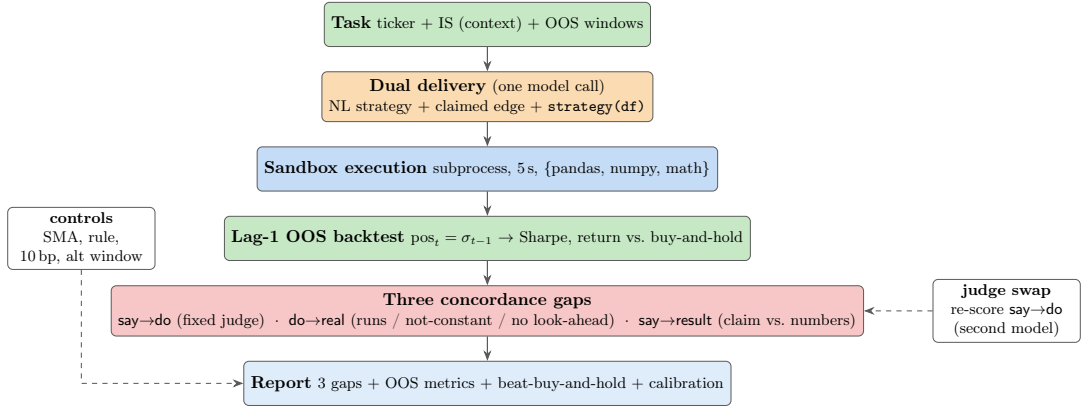
\begin{figure}[t]
  \centering
  \resizebox{0.86\linewidth}{!}{%
  \begin{tikzpicture}[
    font=\small,>={Stealth[length=2.2mm]},
    node distance=5mm and 9mm,
    b/.style={rounded corners=3pt,draw=edgegray,align=center,inner sep=4pt,
              minimum height=9mm,minimum width=52mm},
    s/.style={rounded corners=3pt,draw=edgegray,align=center,inner sep=3pt,
              minimum height=8mm,minimum width=30mm,font=\footnotesize},
    fl/.style={->,draw=edgegray,line width=0.7pt},
    fb/.style={->,draw=edgegray,line width=0.6pt,dashed},
  ]
    \node[b,fill=realcol] (task) {\textbf{Task}\ \footnotesize ticker $+$ IS (context) $+$ OOS windows};
    \node[b,fill=saycol,below=of task] (elicit)
      {\textbf{Dual delivery}\ \footnotesize (one model call)\\
       \footnotesize NL strategy $+$ claimed edge $+$ \strat{}};
    \node[b,fill=docol,below=of elicit] (sbx)
      {\textbf{Sandbox execution}\ \footnotesize subprocess, $5$\,s, \{pandas, numpy, math\}};
    \node[b,fill=realcol,below=of sbx] (bt)
      {\textbf{Lag-1 OOS backtest}\ \footnotesize $\text{pos}_t=\sigma_{t-1}$ $\rightarrow$ Sharpe, return vs.\ \bh{}};
    \node[b,fill=gatered,below=of bt,minimum width=64mm] (gaps)
      {\textbf{Three \concept{} gaps}\\
       \footnotesize \saydo{} (fixed judge) $\;\cdot\;$ \doreal{} (runs / not-constant / no look-ahead) $\;\cdot\;$ \sayresult{} (claim vs.\ numbers)};
    \node[b,fill=docol!55,below=of gaps] (rep)
      {\textbf{Report}\ \footnotesize 3 gaps $+$ OOS metrics $+$ beat-\bh{} $+$ calibration};
    \node[s,fill=white,right=14mm of gaps] (swap) {\textbf{judge swap}\\ \footnotesize re-score \saydo{}\\ \footnotesize (second model)};
    \node[s,fill=white,left=14mm of bt] (ctrl) {\textbf{controls}\\ \footnotesize SMA, rule,\\ \footnotesize 10\,bp, alt window};
    \draw[fl] (task)--(elicit); \draw[fl] (elicit)--(sbx);
    \draw[fl] (sbx)--(bt); \draw[fl] (bt)--(gaps); \draw[fl] (gaps)--(rep);
    \draw[fb] (swap.west)--(gaps.east);
    \draw[fb] (ctrl.south) |- (rep.west);
  \end{tikzpicture}}
  \caption{End-to-end pipeline. One model call yields prose and code; the
  code is executed and backtested on a lag-one OOS engine; the three gaps
  (red hub) reconcile prose, code, and realized outcome. The judge for
  \saydo{} is audited by re-scoring with a second model; SMA, zero-LLM rule,
  cost, and alternate-window controls feed the report.}
  \label{fig:pipeline}
\end{figure}

Code runs in a child process with a wall-clock timeout ($5$\,s) and a
whitelist of imports (\texttt{pandas}, \texttt{numpy}, \texttt{math},
\texttt{statistics}); returned signals are coerced, clipped to
$\{-1,0,1\}$, and reindexed to the bar index. A failed run is recorded as a
row (with $\rho{=}0$), never a crash. \Cref{fig:pipeline} shows the pipeline.

\section{Experimental Setup}
\label{sec:setup}

\paragraph{Universe and windows.}
The universe is eight liquid US ETFs spanning equities, size, international,
bonds, gold, and sectors (SPY, QQQ, IWM, EFA, TLT, GLD, XLF, XLE). The
in-sample window 2018--2022 is provided to the model as context only; the
primary OOS window is 2023--2024, and an alternate 2020--2021 window is used
for regime robustness (\Cref{tab:config}). Data are daily OHLCV on real
tickers.

\paragraph{Models and coverage.}
Four models are run under dual delivery---\texttt{gpt-4o-mini},
\texttt{gpt-4o}, \texttt{gpt-4.1-mini}, and
\texttt{qwen-2.5-7b-instruct}---giving $32$ strategies ($4\times 8$), plus a
two-stage arm on \texttt{gpt-4o-mini} ($8$ more), for $40$ routing units at
seed $42$ and temperature $0.2$. Coverage is deliberately disclosed as a
limitation: Claude, Gemini, DeepSeek, Llama-70B, and Qwen-72B all returned
model-not-found on our inference catalog, so the roster is OpenAI-heavy with
one open model. This is an infrastructure constraint, not a design choice,
and we do not present the leaderboard as a broad model comparison.

\paragraph{Judge and swap.}
\saydo{} is scored by \texttt{gpt-4o-mini}; because that model is also on the
leaderboard, we disclose the conflict and re-score every item with a second
judge, \texttt{qwen-2.5-7b-instruct} (which is in turn circular on the
qwen-7b rows). We report both judges and treat neither as ground truth.

\paragraph{Controls and metrics.}
\Cref{tab:controls} lists the controls. We report the three gaps; OOS Sharpe
and return against \bh{}; the beat-\bh{} rate; position density (fraction of
non-flat days); and calibration of self-reported edge (claimed vs.\ actual
beat-\bh{}).

\begin{table}[t]
  \centering
  \caption{Controls and baselines. \Sys{} measures gaps on the model
  strategies; these reference points isolate whether \concept{} tracks
  anything about performance, and stress the measurements.}
  \label{tab:controls}
  \begin{tabular}{@{}p{0.30\linewidth}p{0.40\linewidth}p{0.22\linewidth}@{}}
    \toprule
    Control & Mechanism & Role \\
    \midrule
    Buy-and-hold        & always long on the OOS slice              & passive benchmark \\
    SMA(50,200)         & long when SMA50 $>$ SMA200, else flat     & non-LLM rule \\
    Rule \saydo{}       & lookback-number overlap, zero LLM         & \saydo{} ceiling \\
    Two-stage           & prose, then code in a second call         & translation control \\
    10 bp cost          & one-way cost on $|\Delta\text{position}|$  & friction robustness \\
    Alt window          & 2020--2021 OOS                            & regime robustness \\
    Judge swap          & re-score \saydo{} with a second model     & judge-robustness audit \\
    \bottomrule
  \end{tabular}
\end{table}

\section{Results}
\label{sec:results}

\subsection{Leaderboard}
\label{sec:leaderboard}
\Cref{tab:leaderboard} reports per-model means. Two patterns appear before
any ablation. The language--code gap is high across the board (\saydo{}
$0.81$--$1.00$) and execution is largely clean (\doreal{} $0.75$--$1.00$),
but \sayresult{} collapses---three of five arms score $0.00$, meaning their
stated edge never survives the numbers. Profit tracks \sayresult{}, not the
other two gaps: every dual model's mean OOS Sharpe is near or below zero,
the best being a marginal \texttt{gpt-4o} at $+0.07$, and none is
competitive with a passive rule (\Cref{sec:concordance}). The two-stage
arm's $0.001$ position density marks it as almost always flat
(\Cref{sec:twostage}), not as a better strategy.

\begin{table}[t]
  \centering
  \caption{Leaderboard (per-model means, $n{=}8$ tasks each). Gaps in
  $[0,1]$; Sharpe is OOS. ``Run fails'' counts strategies that did not
  execute to a valid signal series. The two-stage row is an elicitation
  arm, not a fifth model.}
  \label{tab:leaderboard}
  \begin{tabular}{@{}lcccrcr@{}}
    \toprule
    Model & \saydo{} & \doreal{} & \sayresult{} & OOS Sharpe & pos.\ dens. & fails \\
    \midrule
    \texttt{gpt-4.1-mini}      & \textbf{1.000} & 0.917 & 0.000 & $-0.097$ & 0.290 & 1 \\
    \texttt{gpt-4o}            & 0.938 & \textbf{1.000} & 0.125 & $+0.070$ & 0.639 & 0 \\
    \texttt{gpt-4o-mini}       & 0.844 & 0.917 & 0.000 & $-0.341$ & 0.307 & 1 \\
    \texttt{qwen-2.5-7b}       & 0.812 & 0.917 & 0.000 & $+0.030$ & 0.776 & 1 \\
    \midrule
    \texttt{gpt-4o-mini} (two-stage) & 0.812 & 0.750 & 0.500 & $-0.012$ & 0.001 & 2 \\
    \bottomrule
  \end{tabular}
\end{table}

\subsection{Concordance does not predict alpha}
\label{sec:concordance}
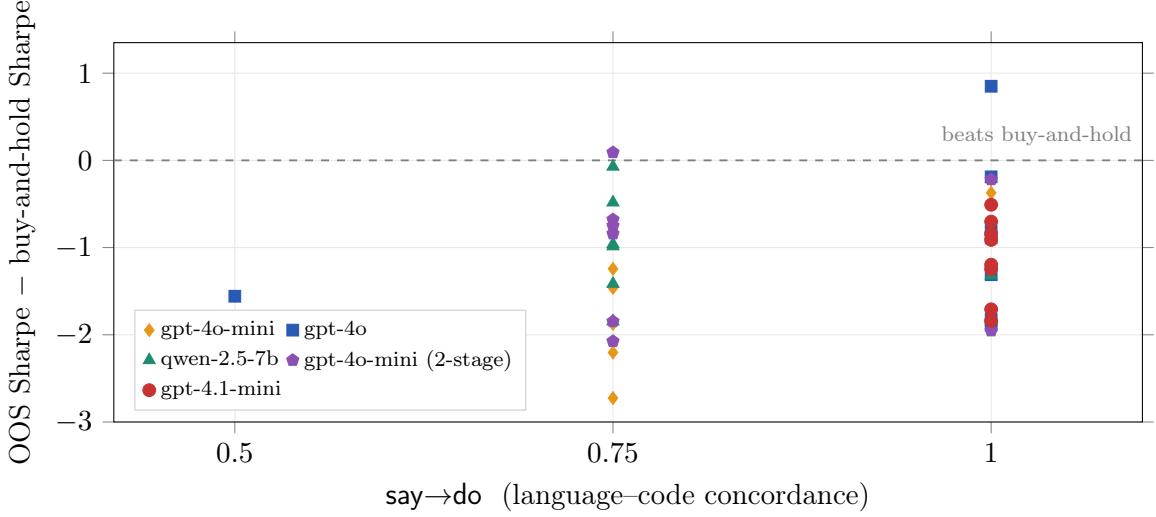
\begin{figure}[t]
  \centering
  \begin{tikzpicture}
  \begin{axis}[
    width=0.92\linewidth, height=6.6cm,
    xlabel={\saydo{} \ (language--code concordance)},
    ylabel={OOS Sharpe $-$ \bh{} Sharpe},
    xmin=0.42, xmax=1.10, ymin=-3.0, ymax=1.35,
    xtick={0.5,0.75,1.0}, ytick={-3,-2,-1,0,1},
    tick align=outside, grid=major, grid style={gray!18},
    legend style={font=\scriptsize, at={(0.02,0.03)}, anchor=south west,
                  cells={anchor=west}, draw=gray!40},
    legend columns=2,
  ]
    \addplot[only marks, mark=diamond*, mark size=2.4pt, color=c4omini]
      coordinates {(0.75,-2.203)(0.75,-1.244)(0.75,-1.462)(1.0,-0.804)(1.0,-0.372)(0.75,-1.874)(0.75,-2.726)(1.0,-0.219)};
    \addlegendentry{gpt-4o-mini}
    \addplot[only marks, mark=square*, mark size=2.2pt, color=c4o]
      coordinates {(1.0,-1.825)(1.0,-1.890)(1.0,-1.311)(1.0,-0.888)(1.0,0.850)(0.5,-1.558)(1.0,-0.189)(1.0,-0.809)};
    \addlegendentry{gpt-4o}
    \addplot[only marks, mark=triangle*, mark size=2.6pt, color=cqwen]
      coordinates {(0.75,-1.847)(0.75,-0.483)(0.75,-1.415)(0.75,-0.986)(0.75,-0.072)(0.75,-0.961)(1.0,-0.854)(1.0,-1.319)};
    \addlegendentry{qwen-2.5-7b}
    \addplot[only marks, mark=pentagon*, mark size=2.4pt, color=c2stage]
      coordinates {(0.75,-1.847)(1.0,-1.953)(0.75,-0.750)(0.75,-0.841)(0.75,0.091)(0.75,-0.678)(0.75,-2.074)(1.0,-0.219)};
    \addlegendentry{gpt-4o-mini (2-stage)}
    \addplot[only marks, mark=*, mark size=2.4pt, color=c41mini]
      coordinates {(1.0,-1.245)(1.0,-1.708)(1.0,-1.840)(1.0,-0.841)(1.0,-0.702)(1.0,-0.912)(1.0,-1.196)(1.0,-0.509)};
    \addlegendentry{gpt-4.1-mini}
    \draw[gray,dashed,line width=0.7pt] (axis cs:0.42,0) -- (axis cs:1.10,0);
    \node[anchor=east,font=\scriptsize,gray] at (axis cs:1.10,0.28) {beats \bh{}};
  \end{axis}
  \end{tikzpicture}
  \caption{\Concept{} does not predict alpha. Each point is one strategy:
  $x$ is its \saydo{} score, $y$ its OOS Sharpe minus \bh{} Sharpe (above
  the dashed line beats \bh{} on Sharpe). The perfect-\saydo{}
  \texttt{gpt-4.1-mini} (red) is entirely below the line; only two of forty
  strategies clear it, both on TLT and one of them a flat degenerate.}
  \label{fig:scatter}
\end{figure}

\Cref{fig:scatter} plots each strategy's \saydo{} against its Sharpe excess
over \bh{}. If speech--code \concept{} bought performance, high-\saydo{}
points would sit above the zero line; they do not. \texttt{gpt-4.1-mini},
the only model with a perfect \saydo{} of $1.00$ on every task, lies
entirely below zero---its most ``faithful'' strategies still lose to holding
the ETF. Across all $40$ strategies only $2$ beat \bh{} on both Sharpe and
return (\Cref{tab:windows}), and the gate makes the non-relationship
explicit: strategies with \saydo{}${=}1.0$ beat \bh{} once in $22$ ($4.5\%$),
those with \saydo{}${<}1.0$ once in $18$ ($5.6\%$)---indistinguishable
(\Cref{tab:gate}). A plain \textsc{sma}(50,200) rule on the same engine
averages $0.583$ OOS Sharpe, above \emph{every} model's mean, and only $8$
of $40$ LLM strategies clear even that rule's Sharpe.

The result survives perturbation: the beat-\bh{} count is $2$ in the
2020--21 window and $2$ after $10$ bp costs (\Cref{tab:windows}), and the
two 2023--24 winners are not the same tasks as the 2020--21 winners. Both
2023--24 winners are on TLT, and one is degenerate---a two-stage strategy
that stayed flat while TLT fell, ``beating'' a declining benchmark without
taking a position (\Cref{app:examples}). \Concept{} is thus necessary
bookkeeping, not evidence of edge.

\begin{table}[t]
  \centering
  \begin{minipage}[t]{0.46\linewidth}
    \centering
    \caption{Gate: does higher \saydo{} beat \bh{} more often? It does not.}
    \label{tab:gate}
    \begin{tabular}{@{}lcc@{}}
      \toprule
      \saydo{} bucket & $n$ & beat \bh{} \\
      \midrule
      $=1.0$ (high) & 22 & 1 \\
      $<1.0$ (low)  & 18 & 1 \\
      \bottomrule
    \end{tabular}
  \end{minipage}\hfill
  \begin{minipage}[t]{0.50\linewidth}
    \centering
    \caption{Beating \bh{} is rare and fragile ($/\,40$).}
    \label{tab:windows}
    \begin{tabular}{@{}lc@{}}
      \toprule
      Criterion & count \\
      \midrule
      beat \bh{}, 2023--24              & 2 \\
      beat \bh{}, 2020--21              & 2 \\
      beat \bh{}, 2023--24 after 10 bp  & 2 \\
      OOS Sharpe $>$ SMA(50,200)        & 8 \\
      \bottomrule
    \end{tabular}
  \end{minipage}
\end{table}

\subsection{Self-assessment is badly calibrated}
\label{sec:calibration}
Models are confident and wrong. Of the $40$ strategies, $32$ explicitly
claim to beat \bh{}; exactly $1$ does---a $3.1\%$ hit rate. This is why
\sayresult{} is $0.00$ for most arms (\Cref{tab:leaderboard}): the gap
records claimed outperformance that the OOS numbers refuse. The lone
two-stage \sayresult{} of $0.50$ is not a counterexample---it is mostly the
``no testable claim detected'' default (\Cref{eq:sayresult}) firing on terse
two-stage rationales, not a detected-and-confirmed edge. A user acting on
these models' stated confidence would have been wrong $31$ times out of
$32$.

\subsection{The language--code judge is not an oracle}
\label{sec:judge}
\saydo{} is the one gap scored by an LLM, so we audit it by swapping the
judge. Under the original judge (\texttt{gpt-4o-mini}) the five arms span
$0.81$--$1.00$; under a Qwen-7B judge they compress to a flat
$0.72$--$0.75$ (\Cref{tab:judgeswap}, \Cref{fig:judgeswap}). The two judges
agree exactly on $16$ of $40$ items, with mean absolute difference $0.15$,
and the model ranking collapses. Two caveats sharpen the reading: the
original judge is itself on the leaderboard, and the swap judge shares a
family with the qwen-7b rows it scores. A zero-LLM rule---counting lookback
numbers common to prose and code---averages $0.975$; the surface tokens
almost always match, so this is a weak ceiling, not a substitute for the
judge, and its near-saturation is itself evidence that \saydo{}'s
discriminating signal lives where a keyword rule cannot see. The lesson is
not that one judge is right, but that a single judge's score is one
measurement: we report both and neither as ground truth.

\begin{table}[t]
  \centering
  \begin{minipage}[t]{0.46\linewidth}
    \centering
    \vspace{0pt}
    \caption{\saydo{} under two judges. Rankings collapse; the swap judge
    compresses to $\approx 0.75$.}
    \label{tab:judgeswap}
    \begin{tabular}{@{}lcc@{}}
      \toprule
      Model & orig & Qwen-7B \\
      \midrule
      \texttt{gpt-4.1-mini} & 1.000 & 0.750 \\
      \texttt{gpt-4o}       & 0.938 & 0.750 \\
      \texttt{gpt-4o-mini}  & 0.844 & 0.750 \\
      two-stage             & 0.812 & 0.750 \\
      \texttt{qwen-2.5-7b}  & 0.812 & 0.719 \\
      \bottomrule
    \end{tabular}
  \end{minipage}\hfill
  \begin{minipage}[t]{0.50\linewidth}
    \centering
    \vspace{0pt}
    \begin{tikzpicture}
    \begin{axis}[
      width=\linewidth, height=5.0cm,
      xlabel={original judge \saydo{}}, ylabel={Qwen-7B \saydo{}},
      xmin=0.4,xmax=1.1, ymin=0.4,ymax=1.1,
      xtick={0.5,0.75,1.0}, ytick={0.5,0.75,1.0},
      tick align=outside, grid=major, grid style={gray!18},
      title={\footnotesize paired \saydo{} ($n{=}40$)}, title style={yshift=-1mm},
    ]
      \addplot[domain=0.4:1.1, samples=2, dashed, gray]{x};
      \addplot[only marks, mark=*, color=c4o, mark size=6.4pt, fill opacity=0.45] coordinates {(1.0,0.75)};
      \addplot[only marks, mark=*, color=cqwen, mark size=5.4pt, fill opacity=0.45] coordinates {(0.75,0.75)};
      \addplot[only marks, mark=*, color=c2stage, mark size=2.0pt] coordinates {(0.5,0.75)};
      \addplot[only marks, mark=*, color=c41mini, mark size=2.0pt] coordinates {(0.75,0.5)};
      \node[anchor=south,font=\scriptsize] at (axis cs:1.0,0.79) {22};
      \node[anchor=south,font=\scriptsize] at (axis cs:0.75,0.79) {16};
    \end{axis}
    \end{tikzpicture}
    
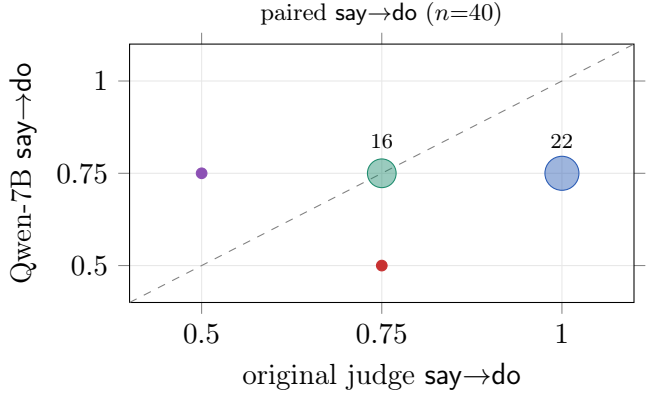
\captionof{figure}{Bubble area $\propto$ count. Points sit on a flat
    $y{\approx}0.75$ band, far from the $y{=}x$ line for high-\saydo{}
    items.}
    \label{fig:judgeswap}
  \end{minipage}
\end{table}

\subsection{Look-ahead dose--response, and an honest negative}
\label{sec:dose}
Two of the three gaps are anchored by execution, so we validate the anchor
rather than assume it. We inject a one-line look-ahead---
\texttt{Close.shift(-1)} at the top of each strategy---and re-score. Over
the $35$ strategies that still run, mean \doreal{} drops by exactly $0.333$
(\Cref{fig:dose}, \Cref{tab:dose}): the static look-ahead check catches the
injected leakage in every case, the intended dose--response.

We also report a negative result plainly. An independent \emph{runtime}
probe---shuffle post-cut prices and test whether in-sample-era signals
move---fired on neither clean nor injected code ($0$ of $40$ both times).
This is a limitation of that probe, not evidence of clean execution: the
shuffle boundary (2022-06-01) and the observation window (through
2021-12-31) do not overlap, so a short-lookback strategy is unaffected by
construction. We keep the static check, drop the runtime claim, and flag the
fix as future work (\Cref{sec:conclusion}).

\begin{table}[t]
  \centering
  \begin{minipage}[t]{0.52\linewidth}
    \centering
    \vspace{0pt}
    \caption{Look-ahead diagnostics. The static check responds to injected
    leakage; the runtime probe does not fire either way (disclosed
    limitation).}
    \label{tab:dose}
    \begin{tabular}{@{}lc@{}}
      \toprule
      Check & value \\
      \midrule
      strategies injected (ran OK)                 & 35 \\
      mean \doreal{} drop after \texttt{shift(-1)} & 0.333 \\
      runtime flag on clean code                   & 0 \\
      runtime flag on injected code                & 0 \\
      \bottomrule
    \end{tabular}
  \end{minipage}\hfill
  \begin{minipage}[t]{0.44\linewidth}
    \centering
    \vspace{0pt}
    \begin{tikzpicture}
    \begin{axis}[
      ybar, width=\linewidth, height=4.8cm,
      title={\footnotesize \doreal{} under injection}, title style={yshift=-1mm},
      symbolic x coords={clean, injected}, xtick=data,
      ymin=0, ymax=1.08, ylabel={mean \doreal{}},
      nodes near coords, nodes near coords style={font=\scriptsize},
      every node near coord/.append style={/pgf/number format/fixed,
        /pgf/number format/precision=2},
      bar width=12pt, enlarge x limits=0.5,
      x tick label style={font=\footnotesize},
    ]
      \addplot[fill=docol, draw=edgegray] coordinates {(clean,0.981) (injected,0.648)};
    \end{axis}
    \end{tikzpicture}
    
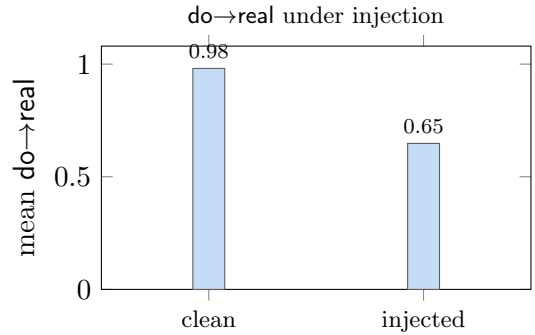
\captionof{figure}{\doreal{} before and after injecting
    \texttt{Close.shift(-1)} (35 running strategies).}
    \label{fig:dose}
  \end{minipage}
\end{table}

\subsection{Dual vs.\ two-stage elicitation}
\label{sec:twostage}
Dual delivery is the default; two-stage (prose, then a separate code call)
controls for the translation step. On the shared \texttt{gpt-4o-mini} base,
dual is modestly higher on both language--code gaps (\saydo{} $0.844$ vs.\
$0.812$; \doreal{} $0.917$ vs.\ $0.750$). Two-stage looks less bad on Sharpe
($-0.012$ vs.\ $-0.341$) only because its strategies are almost always
flat---position density $0.001$ versus $0.307$---so it neither loses nor
wins. Staying out of the market is not a strategy; the \concept{} gaps, not
the Sharpe, are the honest comparison.

\section{Discussion}
\label{sec:discussion}

\paragraph{Implications.}
The headline is a separation. An LLM can be highly self-consistent---its
code faithfully implements its prose---and still produce nothing that beats
holding the asset. \Concept{} and competence are different axes, and
conflating them, as a fluent, well-coded, confidently-captioned strategy
invites, is exactly the error a procurement or deployment decision cannot
afford. Two of our three gaps can be measured cheaply and reliably from
execution; the third---self-reported edge---is the one to distrust, and it
is both the most confident and the least calibrated signal in the system.

\paragraph{Practitioner guidance.}
Read the gaps separately, not as an average. \doreal{} (does it run, without
look-ahead) is a hard gate worth trusting. \sayresult{} (does the claim
hold) should be treated as a claim to be checked against a non-gameable
backtest, never taken at face value. \saydo{} should be scored by a fixed
judge that is not among the models under test, and reported alongside a
swapped-judge number so readers can see its fragility. Profit should never
be optimized inside the loop: repairing strategies to raise Sharpe destroys
the identification the protocol provides.

\paragraph{Limitations.}
Several bound our claims. The sample is small ($40$ strategies, single seed
at temperature $0.2$), so we report effects, not confidence intervals. Model
coverage is OpenAI-heavy with one open model, because Claude, Gemini,
DeepSeek, Llama-70B, and Qwen-72B all returned not-found on our inference
catalog---an infrastructure limitation, but a real one. The universe is
eight liquid US ETFs and rule-based daily strategies only; we do not test
intraday, multi-asset, portfolio, or learned strategies. The \saydo{} judge
is an LLM (audited but not resolved), the original judge sits on the
leaderboard, and the swap judge is circular on its own family's rows.
\sayresult{} uses a keyword parser that returns ``undetermined'' when it
cannot match a claim. Most importantly, \Sys{} measures internal \concept{}
and execution fidelity, not tradability: a concordant, honestly-captioned
strategy is not thereby a good one, and nothing here is investment advice or
a claim that any model beats the market.

\section{Conclusion and Future Work}
\label{sec:conclusion}
We presented \Sys{}, a protocol that treats an LLM trading strategy as three
artifacts to reconcile---the stated logic, the generated code, and the
realized track record---and scores three \concept{} gaps anchored by a
backtest the model cannot game. Across $8$ ETFs and four models, \concept{}
did not predict profit, self-reported edge was badly calibrated ($1$ of $32$
claims held), and the one LLM-scored gap was not robust to swapping the
judge; a look-ahead dose--response validated the execution gap while we
disclosed a runtime probe that did not work. We frame these as measurement
results and delimit external validity as future work.

The clearest next steps follow the limitations: broaden the model roster
once catalog access allows and add seeds for confidence intervals; replace
the keyword \sayresult{} parser with a claim extractor; fix the runtime
look-ahead probe so its shuffle boundary overlaps the observation window, or
replace it with a stronger dynamic test; and extend beyond single-asset rule
strategies toward portfolios and a genuine forward test, so that \concept{}
can eventually be related to realized, out-of-sample tradability rather than
to a closed backtest.

\clearpage
\bibliographystyle{plainnat}
\bibliography{references}

\appendix
\begin{center}
  {\Large\bfseries Appendix}
\end{center}
\vspace{0.5em}
\section{Configuration}
\label{app:config}
\begin{table}[H]
  \centering
  \caption{Frozen v0 configuration.}
  \label{tab:config}
  \begin{tabular}{@{}ll@{}}
    \toprule
    Parameter & Value \\
    \midrule
    universe                    & SPY, QQQ, IWM, EFA, TLT, GLD, XLF, XLE \\
    in-sample (context only)    & 2018-01-01 \,--\, 2022-12-31 \\
    out-of-sample (primary)     & 2023-01-01 \,--\, 2024-12-31 \\
    alt window (ablation)       & 2020-01-01 \,--\, 2021-12-31 \\
    dual models                 & gpt-4o-mini, gpt-4o, gpt-4.1-mini, qwen-2.5-7b-instruct \\
    two-stage arm               & gpt-4o-mini (elicitation ablation) \\
    routing units               & 40 \ (32 dual $+$ 8 two-stage) \\
    seed / samples / temperature& 42 \,/\, 1 \,/\, 0.2 \\
    backtest                    & $\text{pos}_t=\sigma_{t-1}$;\ Sharpe $\times\sqrt{252}$ \\
    costs                       & 0 bp (main);\ 10 bp one-way (ablation) \\
    SMA baseline                & 50 / 200, long-only \\
    sandbox                     & subprocess, 5\,s, \{pandas, numpy, math, statistics\} \\
    \saydo{} judge              & gpt-4o-mini (orig);\ qwen-2.5-7b-instruct (swap) \\
    lookback set (rule \saydo{})& \{5, 10, 14, 20, 21, 50, 100, 200\} \\
    \bottomrule
  \end{tabular}
\end{table}

\section{The two ``winners''}
\label{app:examples}
Only two of the forty strategies beat \bh{} on the 2023--24 window, both on
TLT, and they illustrate opposite failure modes of reading a raw beat-\bh{}
count.

\emph{Real-ish (\texttt{gpt-4o} / TLT).} An RSI mean-reversion rule with a
$20$-day moving-average filter and a volume liquidity gate; claimed edge:
that TLT ``exhibits short-term mean-reversion \dots\ which should outperform
a buy-and-hold.'' It realized OOS Sharpe $0.759$ vs.\ \bh{} $-0.091$ and
return $+0.084$ vs.\ $-0.055$, at position density $0.098$---a genuine, if
sparse, edge on this one ticker and window.

\emph{Degenerate (\texttt{gpt-4o-mini} two-stage / TLT).} A strategy that
resolved to staying flat: zero position, zero return, and (correctly) no
claim to beat \bh{}, so its \sayresult{} is undetermined rather than
confirmed. It ``beats'' \bh{} only because TLT fell over the window and
holding cash did not. Staying out of a falling asset is not a strategy; this
row is why the calibration count of claimed-and-confirmed wins is $1$, not
$2$.

\end{document}